\documentclass{article} 
\usepackage{graphicx,amssymb} 
\usepackage{bm}

\begin{document} 

\title{From liquid to solid bonding in cohesive granular media} 
\author{Jean-Yves Delenne\footnote{Corresponding author: delenne@lmgc.univ-montp2.fr; Laboratoire de M\'ecanique et G\'enie Civil,  CNRS - Universit\'e Montpellier 2, Place Eug\`ene Bataillon, 34095 Montpellier cedex 05}, Fabien Souli\'e, Moulay Sa\"id El Youssoufi and Farhang Radjai}

\maketitle

\begin{abstract} 
We study the transition of a granular packing  from liquid to 
solid bonding in the course of drying. The particles are  
initially wetted by a liquid brine and the cohesion of the packing is ensured by capillary forces, 
but the crystallization of the solute transforms the liquid bonds into partially cemented 
bonds. This transition is evidenced experimentally by measuring the 
compressive strength of the samples at regular intervals of  times. 
Our experimental data reveal three regimes: 
1) Up to a critical degree of saturation, no solid bonds are formed and 
the cohesion remains practically constant; 2) The onset of cementation 
occurs at the surface and a front spreads towards the center of the sample with 
a nonlinear increase of the cohesion; 3) All bonds are partially cemented 
when the cementation front reaches the center of the sample, but the cohesion increases 
rapidly due to the consolidation of cemented bonds. We introduce a model based 
on a parametric cohesion law at the bonds and a bond crystallization parameter. This model 
predicts correctly the phase transition and the relation between microscopic 
and macroscopic cohesion.

\noindent keywords: Unsaturated granular media, capillarity, cementation, DEM, compressive strength

\end{abstract}

\section{Introduction}
\label{sec:intro}

Granular materials occur very often in nature and industrial applications 
with liquid or solid bonds between the particles.   
The effect of capillary bonding on the mechanical behavior of 
granular materials is 
of primary importance in powder technology \cite{Iveson96,Iveson01,Nokhodchi05} 
and transformations of geomaterials \cite{Cardell03,Lubelli04,Rijniers05}.  
Capillary bonding has 
been extensively studied in the past, and several 
models of capillary cohesion have been 
proposed \cite{Mikami98,Willett00,Soulie06}.  
The scale-up of capillary forces to  
the macroscopic cohesion of granular media has been 
investigated by means of both experiments 
and discrete element numerical simulations  \cite{Pepin01,Kim03,Richefeu06}.  

In the same way, solid bonding is found in various materials such as  
sedimentary rocks (sandstones, conglomerates and breccia)\cite{Tarbuck2005}, 
biomaterials such as wheat  endosperm (starch granules forming a compact structure 
bound together by a protein matrix) \cite{Topin2008,Topin2008a}, and 
geomaterials like mortars, concrete and asphalt 
(aggregates of various sizes glued to each other by a cement paste) \cite{Schlangen1992}. 
The cohesive behavior due to solid bonds between the particles 
and the rupture of cemented 
granular materials have been recently investigated by several authors 
\cite{Delenne04, Delenne08,Topin2007b}. 

There is a broad class of processes where the nature of cohesive bonds 
evolves from solid to liquid and vice versa.     
The liquid is generally water with  various types of impurities and/or dissolved minerals. 
As a result of reactive transfers or phase transitions, the 
liquid may influence the mechanical properties by dissolution of cemented 
joints or crystallization of solutes. 
The mechanical properties can thus evolve in response to 
external hygrothermic conditions \cite{Soulie06}. 
In pharmaceutical industry, for example, the process of wet granulation is 
widely used to manufacture tablets.  A liquid  
composed of solvent(s) and solute(s) is mixed with dry powder  
in order to obtain a homogeneous mixture that would segregate 
in the absence of the liquid.   The formation of aggregates 
in the process of mixing has been a subject of extensive  
investigation \cite{Iveson96,Iveson01,Nokhodchi05}. 
The solid tablets are obtained by drying:  the solvent evaporates 
and solid bridges are formed between grains by crystallization, thus ensuring the macroscopic 
cohesion of the tablets. 

In this paper, we are interested in the evolution of 
the macroscopic cohesion of a wet granular material during the 
evaporation of a solvent (water) and thus crystallization 
of the solute (sodium chloride). 
At low liquid contents, the water is in the form of capillary bridges between particles.  
The hydrous field corresponding to this range of liquid contents is 
often described as ``pendular''. The attractive capillary forces in the 
presence of liquid bridges between 
particles endows the material with macroscopic cohesion \cite{Soulie06,Richefeu2006a}.  
Under the action of the surrounding 
hygrothermic conditions, the water  evaporates and 
the solutes crystallize to form solid  bonds between grains.  
The cohesion is initially of capillary origin but begins to 
increase as soon as the first solid bonds are formed at the 
free boundary of the sample and a solidification 
front  propagates from the boundary to the center 
of the sample. 

We measure the compressive strength 
of several samples of glass beads and sand, prepared in the pendular state 
by mixing with a saturated brine, in the course of drying. 
Technically, it is not trivial to follow simultaneously the evolution of 
the internal state due to the evaporation of the brine and the evolution of its 
mechanical behavior. Moreover, the two limits of purely liquid bonding 
and purely solid bonding have different behaviors and require specific 
experimental tools and precautions. For these reasons, the experimental 
campaign involves the preparation of many reproducible 
samples subjected to identical drying conditions, each sample being  
destroyed when subjected to simple compression test.   

We show that the evolution of the cohesion  is governed by 
three time scales corresponding to the evaporation of a single 
capillary bridge, the propagation of the cementation front 
from the surface to the center and 
the aging of cemented bonds. These time scales are identified 
from the experimental campaign together with a theoretical model based on a 
a simple representation of the evolution of 
the debonding force between the particles. This model  
fits well the compressive strength as a function of an order 
 parameter representing the state of crystallization.  

In the following, we first describe in section \ref{sec:exp} 
the experimental setup, the 
procedures of sample preparation and the main results regarding the evolution 
of the samples and mechanical strength in the course of drying. 
In section \ref{sec:model}, we present our model 
and compare its predictions with the experimental data. 
We conclude with a summary and scopes of this work. 
 
\section{Experiments}
\label{sec:exp}
\subsection{Experimental setup}

The experimental tests were designed to measure the compressive strength 
of granular samples during the process of the drying and deposition of a solute 
at the liquid bridges between the grains.    
Two types of materials were used in these tests: 
(1) a model material composed of glass beads and (2) 
a natural material: Ventoux sand. 
In both cases, the grains were first washed and dried. 
Then, they were sieved to keep only the grains with
diameters in the range from 0.4 mm to 0.8 mm. 
The dry grains were homogeneously mixed with a small amount of saturated 
brine containing  35.6 g of NaCl for 100 g of water in controlled environment (temperature, humidity).
This maximum concentration varies with  temperature \cite{Kaufmann60,Langer82}. 

The samples (diameter 25 mm and height 17 mm) were prepared 
inside a cylindrical mould. 
At this stage, the samples are particularly 
delicate to handle since their mechanical strength 
is solely ensured by weak capillary forces acting between the grains.  
Because of this brittleness, particular experimental precautions were necessary 
in order to avoid damage to the samples upon unmoulding. 

Each sample is initially a three-phase 
material including 
\begin{itemize} 
\item a solid phase (denoted by S) composed of grains (glass beads or sand) 
with a total mass $m^{S}_{grains}$,
\item a liquid phase (denoted by L) composed of water of mass $m^{L}_{H2O}$ 
and dissolved NaCl of mass $m^{L}_{NaCl}$.
\item a gaseous phase composed of air and water vapor with negligible 
mass compared to the two other phases but which carries the kinetics of drying.
\end{itemize}
Thus, the total mass of the sample at this initial state is 
$m_i =  m^L_{H2O}+m^L_{NaCl}+m^S_{grains}$. 

The water evaporates at the surface and causes thermodynamic 
disequilibrium of the solution.  
The equilibrium  is recovered  through the crystallization of 
NaCl and thus the formation of solid NaCl deposits 
of a total mass $m^{S}_{NaCl}$. A crystallization front 
propagates thus from the boundary towards the center 
of the sample.  
The evolution of a sample is expected to be governed by the ratio 
\begin{equation}
I_c = \frac{m^S_{NaCl}}{m^S_{NaCl}+m^L_{NaCl}}
\label{eqn:Ic}
\end{equation} 
to which we refer as the global crystallization index of the sample.  

\begin{figure}
	(a)
	\includegraphics[width=.4\columnwidth]{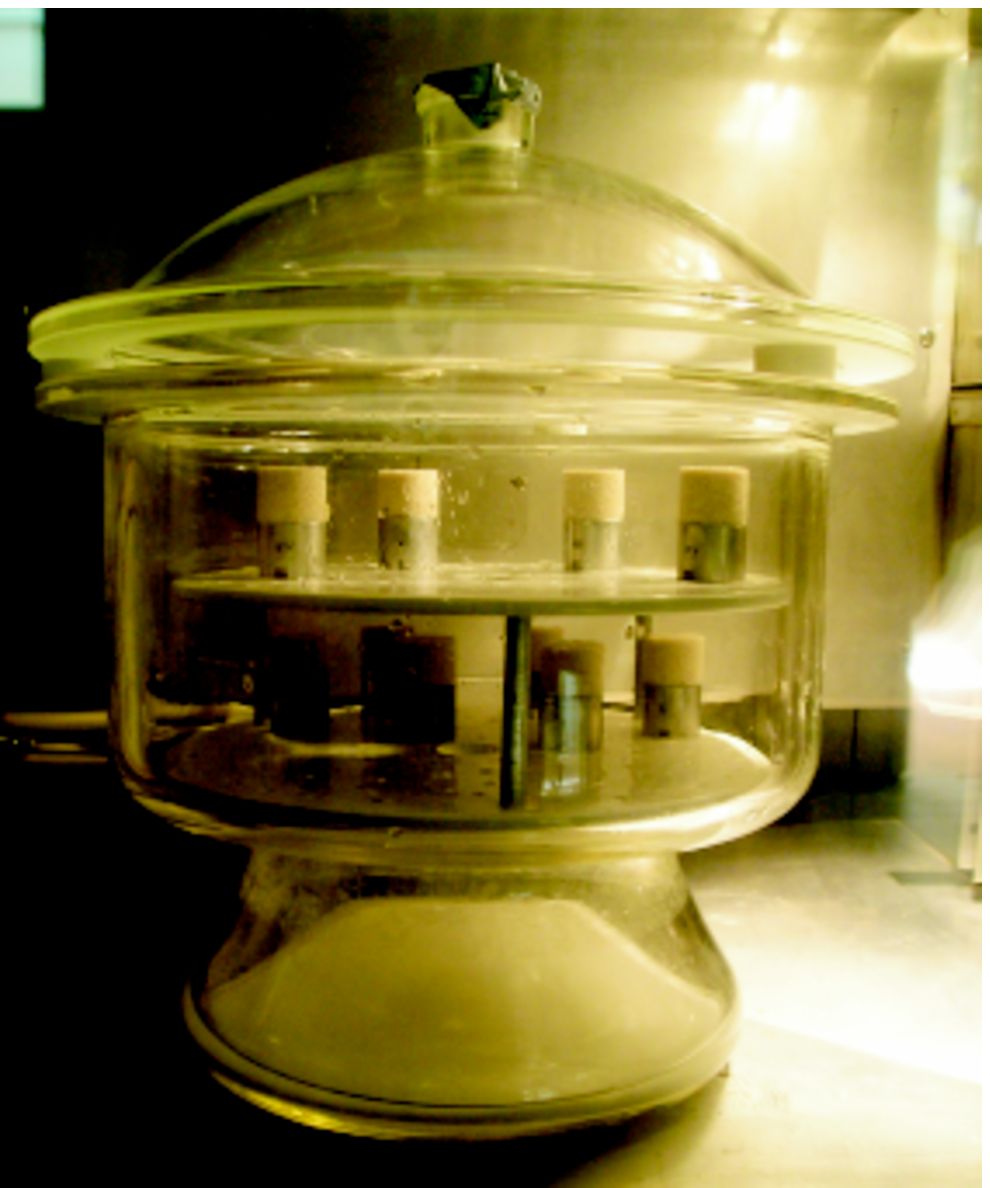}
	(b)
	\includegraphics[width=.4\columnwidth]{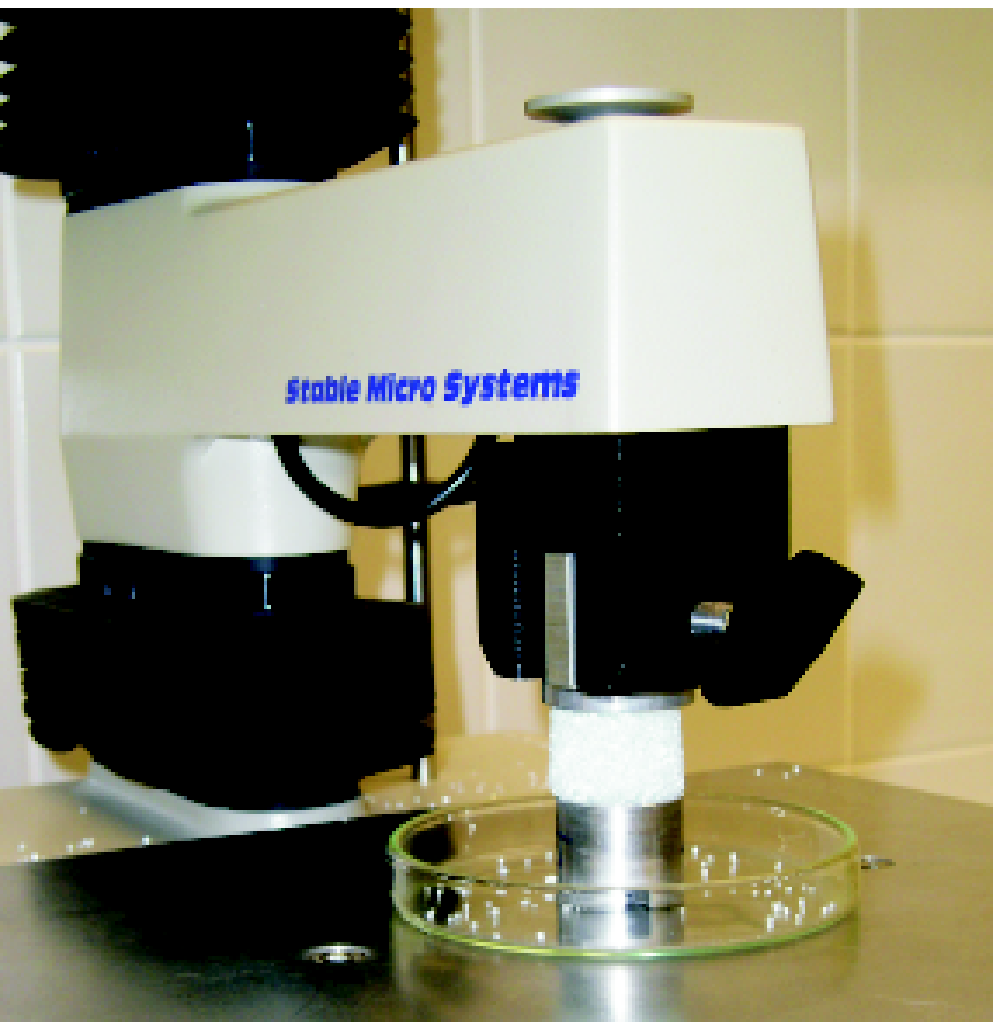}
	\caption{(a) Photograph of the sample in the process of drying; (b) The press used for the unconfined compression tests.}
	\label{fig:Photo_EtuveTexturometre}
\end{figure}

Three series of samples were prepared with three different values   
of the initial liquid content $w_{Li} \equiv (m^L_{H2O}+m^L_{NaCl})/m^S_{grains}$:  
$3\, \%$, $5\, \%$ and $7\, \%$.  
These values were selected in the hydrous field corresponding to 
the ``pendular state'' where liquid bridges between grains 
ensure the mechanical integrity of the sample.
The samples were weighed just after preparation, and left to 
dry at constant  temperature $T=20^\circ$ C and relative humidity 
$RH=43\;\%$ 
for drying times varying from 15 minutes to 18 hours; see Fig. \ref{fig:Photo_EtuveTexturometre}(a).

At regular intervals of time, the samples were  weighed (mass $m_{f}$), and then 
subjected to unconfined vertical compression up to failure.  
The tests were carried out by means of 
a ``low capacity'' press, which allows us to test samples with low dimensions 
and  forces up to 50 N with an accuracy of 0.01 N; see Fig. \ref{fig:Photo_EtuveTexturometre}(b). 
The difference $m_{f}-m_{i}$ yields the mass of 
the evaporated water and thus that of the crystallized mass fraction of NaCl. 
Since all samples are prepared under similar conditions, 
they are assumed to follow the same evolution in time. This allows us   
to reconstruct the kinetics of crystallization.  

\subsection{Experimental results}
\label{sec:experiments}

\begin{figure}
\centerline{ \includegraphics[width=.8\columnwidth]{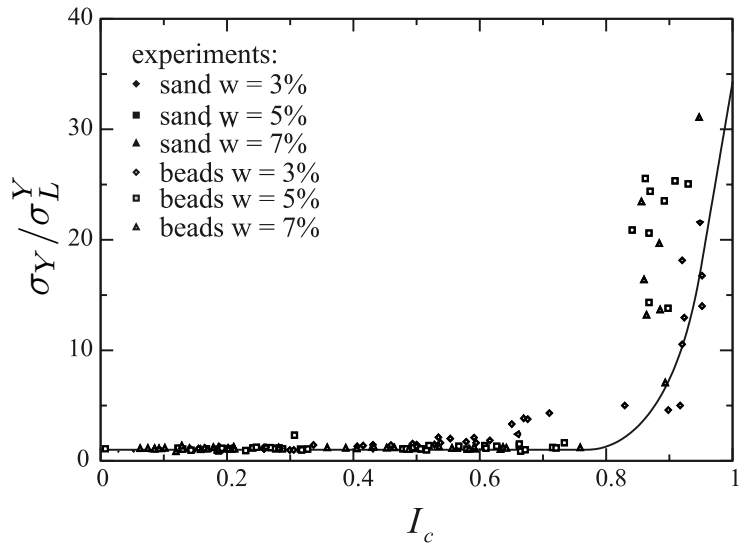}}
	\caption{Compressive yield stress $\sigma^Y$, normalized by the yield stress $\sigma^Y_L$ 
	due only to capillary forces (yield stress in the first regime) 
	as a function of the global crystallization 
	index $I_c$ in experiments. The solid line corresponding to theoretical model (see section 3) is plotted as a guide to eyes.}
	\label{fig:sigmaYIc_exp}
\end{figure}

The evolution of the compressive stress $\sigma^Y$ at yield is shown in 
Fig. \ref{fig:sigmaYIc_exp}  as a function of  the global crystallization index $I_c$ 
for glass beads and sand, as well as from a theoretical model described 
in section \ref{sec:model}.  
All the experimental data  collapse nearly on the same curve 
for glass beads and sand.  
We distinguish three regimes in the evolution of the cohesion.  
As long as $I_c$ is below a threshold $I_T$, the yield stress is independent of $I_c$. 
In this regime, the amount of the crystallized  salt at the 
boundary layer of the sample, where the whole process 
is initiated due to exchange with the atmosphere, 
is insufficient to allow for the formation of solid bonds between the particles. 
Therefore, the yield stress $\sigma^Y$  in this regime is only due to the capillary 
bonds between the particles. In Fig. \ref{fig:sigmaYIc_exp}, we see that $\sigma^Y$ is independent 
both of the initial liquid content $w_{Li}$ and of the type of grains. In fact, both experiments and 
theoretical analysis suggest that in a 
homogeneously wetted granular material in the pendular state, the capillary cohesion 
is practically independent of the liquid content but crucially depends on the number density of 
the capillary bonds \cite{Richefeu2006a}.    

\begin{figure}
\centerline{ \includegraphics[width=.4\columnwidth]{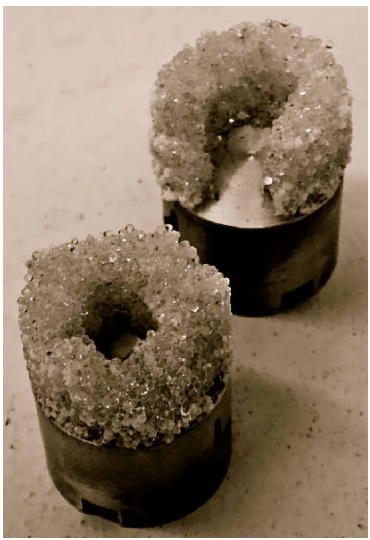}}
	\caption{Photograph of a partially dried sample in the second regime.}
	\label{fig:cementationFront}
\end{figure}

\begin{figure}
\centerline{   \includegraphics[width=.8\columnwidth]{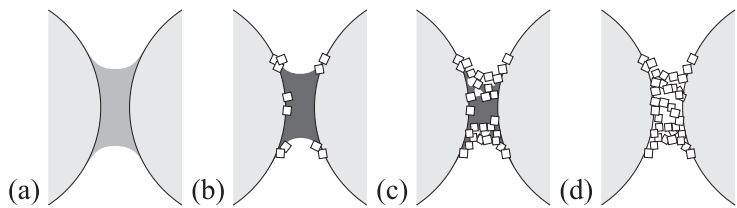}}
  \caption{Schematic representation of the evolution of 
  bridge: from a liquid to a solid bridge.
  \label{fig:Deposition}}
\end{figure}

The yield stress increases from $I_c \simeq 0.9$ 
due to the partial cementation of the first bonds at the boundary of the sample and  the 
propagation of a 
solidification front towards the center of the sample. The yield stress 
increases rapidly in this regime. 
A photograph of the cementation front in a partially dried sample  
is displayed in Fig. \ref{fig:cementationFront}. The 
bonds in the central part of the sample were of capillary type and thus the grains  
were easily removed. The remaining grains form a solid crown due to cemented bridges 
between the particles. Hence, the thickness of this crown represents the distance run by the 
solidification front. 
It should be noted that the solid bonds behind this front are only 
partially cemented. At this stage, each solid bond is composed of a crystallized superficial layer  
with liquid brine inside. As drying continues, 
the cemented bond gets thicker and stronger until the whole bond solidifies. 
This process is illustrated in Fig. \ref{fig:Deposition}. 

All bonds are cemented when the center  of the sample is 
reached by the cementation front. In this third regime,  
$\sigma^Y$ rises linearly to a level as high as 35 times the capillary 
yield stress due to the consolidation of the cemented bonds throughout the system 
as a result of continued crystallization of the brine at each bond. 
These high values of compressive strength  
as $I_c \rightarrow 1$ has two different origins. On one hand, 
the debonding force of a solid NaCl bond 
is nearly 7 times that of a capillary bond. On the other hand, 
unlike a liquid bond which involves only a tensile strength in addition to 
sliding friction, a solid bond has a rolling (or bending) strength 
as well as a tensile strength. 
This rolling strength is crucial for the compressive strength of cohesive 
granular materials \cite{Estrada2008,Taboada06}.   

\begin{figure}
\centerline{\includegraphics[width=0.8\columnwidth]{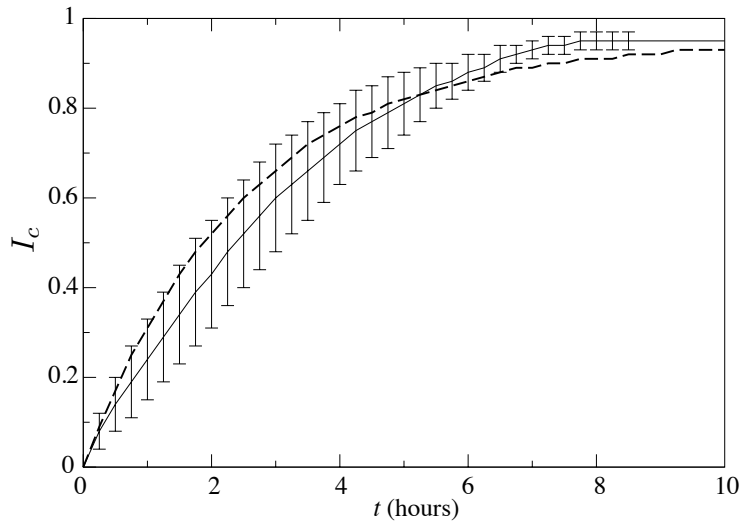}}
  \caption{Evolution of the global crystallization index with time; solid line: experimental results,
  dotted line: model equation (\ref{eqn:ict}). The error bars represent the standard deviation 
  for five independent experiments. 
  \label{fig:Ict_exp}}
\end{figure}

Figure \ref{fig:Ict_exp} shows the evolution of $I_c$ with time $t$. This evolution  
reflects the kinetics of drying with the transport of water in the gas phase.  
$I_c$ begins to increase as soon as 
the sample is exposed, but the first partially cemented  bonds are completely formed only at 
$I_c = I_T \simeq 0.9$. Then, 
$I_c$ increases mainly due to the formation of the first (partially) cemented 
bonds followed by the propagation of 
the cementation front into the sample. When this front reaches the center of 
the sample, $I_c$ increases only due to the aging and consolidation of the cemented
bonds. Although the kinetics of drying is not exactly the same in each regime, 
the time evolution of $I_c$ is quite well fit by an exponentially increasing function:
\begin{equation}
I_c(t) = 1 - e^{-\frac{t}{t_d}}
\label{eqn:ict}
\end{equation} 
where $t_d$ is a characteristic drying time. This form suggests that the drying 
rate is proportional to the residual liquid. 

\section{Model}
\label{sec:model}

The evolution of the compressive strength may be understood on quantitative 
grounds by considering 1) a local cohesion law involving 
the phase change from liquid to solid at the 
level of a single bond,  2) the relation between the bond strength and the macroscopic 
strength, and 3) the kinetics of drying both at the surface and in the 
bulk. In the following, we present a model that is based on these ingredients and 
correctly captures the macroscopic evolution as shown by the fitting form 
in Fig. \ref{fig:sigmaYIc_exp}.   

\subsection{Single-bond behavior}
\begin{figure}[bt]
\centerline{\includegraphics[width=.5\columnwidth]{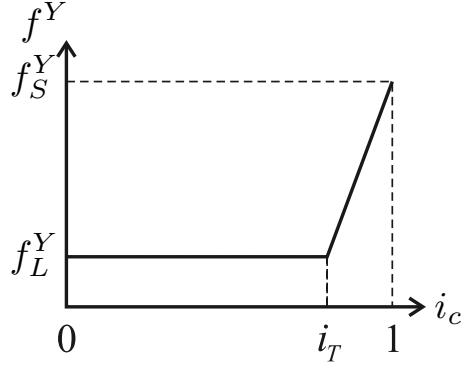}}
  \caption{The parametric adhesion law.}  
  \label{fig:parametricCohesion}
\end{figure}

The phase change of a bond from liquid to solid is 
controlled by the amount of the deposited solute. 
Hence, the natural control parameter for phase 
transition at this scale is the bond crystallization index $i_c$ 
defined by 
\begin{equation}
i_c = \frac{m^S_{NaCl}(\mbox{bond})}{m^S_{NaCl}(\mbox{bond})+m^L_{NaCl}(\mbox{bond})}
\label{eqn:ic}
\end{equation} 
This index depends on the history of the 
bond and is thus a function $i_c(r,t)$ of the radial position $r$ of the bond and 
the time $t$ elapsed since the beginning of drying. We assume translational 
invariance along the axis of the sample. 
The index $i_c$  varies between $0$ for a liquid bond and $1$ for a fully cemented bond. 
The transition from liquid to solid takes place 
at a particular value $i_c = i_T$ which corresponds to the deposition of the first 
percolating shell of crystallized solute across the gap or at contact between two particles. 
In this transition, the tensile force threshold $f^Y$ (the debonding force) increases from that of 
a liquid bond $f^Y_L$ to that of a fully cemented bond $f^Y_S$. We 
have $f^Y = f^Y_L$ for $0 \leq i_c \leq i_T$ and  $f^Y = f^Y_S$ for $i_c=1$. 
Assuming a linear evolution between these two limits, 
we get the following {\em parametric}  adhesion law:
\begin{equation}
f^Y=f^Y_{L}+H\left( i_c - i_{T}\right)  \frac{i_c-i_{T}}{1-i_{T}} (f^Y_S - f^Y_L) 
\label{eqn:loi_evolutive}
\end{equation}
where $H$ is the Heaviside step function. A plot of this relation is shown   in 
Fig. \ref{fig:parametricCohesion}. 

Remark that $f^Y$ in equation \ref{eqn:loi_evolutive} is the force threshold and 
the actual value of the normal force $f_n$ depends on the force law expressing 
a force-displacement relation. This relation is different for a liquid bond and a 
partially or fully cemented bond. Several expressions have been proposed for 
the capillary force $f_L$ as a function of the gap $\delta$, bond liquid volume $V_b$ and 
surface tension $\gamma$ \cite{Soulie06, Richefeu2006a}. 
The capillary force has its largest value at the 
contact between two particles and declines exponentially with the gap: 
\begin{equation} 
f_L= 2\pi \gamma \cos \theta \,\rho \,e^{-\delta / \lambda},
\label{eqn:fnc} 
\end{equation} 
where $\lambda$ is a characteristic length depending on 
$V_b$ and the particle radii, $\rho$ is a reduced length depending on the particle radii 
and $\theta$ is the contact angle \cite{Willett00,Bocquet02,Herminghaus05}. 
A capillary bridge is stable as long as $\delta < \delta^{max}$ given by \cite{Lian98} 
\begin{equation} \label{eq:debond-dist} 
\delta^{max} = \left ( 1 + \frac{\theta}{2} \right ) V_b^{1/3}. 
\end{equation} 
This distance is small (compared to $\lambda$) in the pendular state and in practice the debonding force 
can be approximated by 
\begin{equation} 
f^Y_L= 2\pi \gamma \cos \theta \,\rho 
\label{eqn:fYL} 
\end{equation} 
This force is independent of the bond liquid volume $V_b$.

The fully cemented bonds have a brittle behavior with a debonding 
threshold $f^Y_S\simeq 7 f^Y_L$ for the three values of $w_{Li}$ used 
in our experiments. The behavior is also brittle for the partially cemented bonds 
with a threshold which is assumed here to vary linearly with the bond crystallization index 
according to equation \ref{eqn:loi_evolutive}. Obviously, the cemented 
bonds exhibit also a shear strength and a rolling strength in addition to 
the tensile strength considered above. This feature, which differentiates  
cemented bonds from liquid bonds, is not considered explicitly 
in the single-bond behavior but will be taken into account through a 
Coulomb cohesion parameter $c$ at the macroscopic scale.  

\subsection{From single bond to RVE}

\begin{figure}
\centerline{\includegraphics[width=.5\columnwidth]{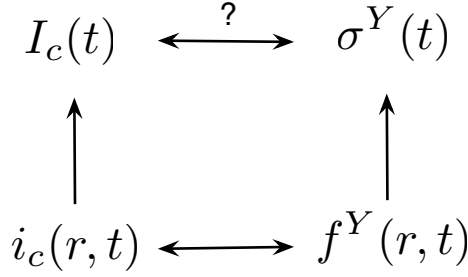}}
  \caption{Diagram of  microscopic model and upscaling method.  
  \label{fig:diagram_mod}}
\end{figure}

A schematic representation of our model is shown in Fig. \ref{fig:diagram_mod}. 
It consists of two scales: 1) The microscopic or local scale is defined by the 
parametric adhesion law $f^Y\{ i_c(r,t) \}$; 2) The macroscopic or global scale 
is characterized by the relation $\sigma^Y\{I_c(t)\}$. Given the local adhesion law, the 
macroscopic behavior can be determined from the relationships between 
$i_c$ and $I_c$, on one hand, and between $f^Y$ and $\sigma^Y$, on the other hand. 
We first consider a Representative Volume Element (RVE). Then, solve the problem 
in the presence of hydric gradients in the drying process by introducing a 
simple propagation equation for the local crystallization index. 

We assume a homogeneous distribution of the liquid bonds in the 
initial preparation of the sample. We also assume that the  total mass of NaCl 
 is the same for all bonds. 
Then, the macroscopic crystallization index is simply the volume average 
of the bond crystallization index:
\begin{equation}
I_c (t)= \langle i_c(r,t) \rangle_V = \frac{h}{V} \int\limits_0^R 2\pi r i_c(r,t) \ dr 
\label{eqn:Icic}
\end{equation}
where $h$ and $R$ are the height and radius of the sample, respectively, 
and $V=\pi R^2 h$ is its volume. According to equation \ref{eqn:Icic}, for a RVE and in the 
absence of hydric gradients, we have $I_T = i_T$ since all bonds reach simultaneously this point. 
During drying, the cementation is initiated at the surface of the sample for $i_c=i_T$   
and we have $i_c<i_T$ everywhere in the bulk. 

In order to propose  a relation between  $f^Y$ and $\sigma^Y$, we 
need to postulate the deformation mechanism, the macroscopic failure 
mode and the general expression of the stress tensor in a granular material 
as a function of the bond forces. The stress tensor $\sigma_{ij}$ is the first moment 
of the bond forces $f_i$ averaged over the control 
volume $V$ \cite{Taboada06,Voivret09}:
\begin{equation}
\sigma_{ij} = n_b \langle f_i \ell_j \rangle_V
\end{equation} 
where $n_b$ is the number density of the bonds and $\ell_j$ is the $j$-component 
of the branch vector joining the centers of the two particles at contact. The averaging  
runs over all contacts belonging to the control volume. Assuming that the branch vector lengths 
and forces are not correlated and for a medium with uniformly distributed branch vectors in all 
directions, the normal stress $\sigma_n$  is given by 
\begin{equation}
\sigma_n = \frac{1}{3} n_b \langle \ell \rangle \langle f_n \rangle
\label{eqn:sigman}
\end{equation} 
where $f_n$ is the normal bond force. We have $\langle f_t \rangle \simeq 0$ 
as a result of the balance of force moments on each particle. 

According to equation \ref{eqn:sigman}, the maximum tensile force 
$\sigma_n^{max}$ in a RVE is obtained by replacing $f_n$ by $f^Y$:
 \begin{equation}
\sigma_n^{max} = \frac{1}{3} n_b \langle \ell \rangle \langle f^Y \rangle
\label{eqn:sigmanmax}
\end{equation} 
Assuming that the Mohr-Coulomb failure criterion holds in the 
tensile regime, the Coulomb cohesion $c$ is thus simply given by 
\begin{equation}
c= \mu \sigma_n^{max} = \frac{1}{3} n_b \tan \varphi  \langle \ell \rangle \langle f^Y \rangle
\label{eqn:c}
\end{equation}
where $\mu$ is the friction coefficient and $\varphi$ is the friction angle.

Finally,  we  need  the relation between the axial stress $\sigma_1$ 
and the cohesion $c$ in order to evaluate the compressive strength. 
Under the conditions of axial symmetry, we have $\sigma_2=\sigma_3$. 
Let $p=(\sigma_1 + 2\sigma_3)/3$ and 
$q=(\sigma_1 - \sigma_3)/3$ be the mean stress and stress deviator, respectively. 
At failure, we have 
\begin{equation}
\frac{q}{p} = \frac{2}{3-\sin \varphi} \left( \sin \varphi + \frac{c}{p} \cos \varphi \right)
\label{eqn:qp}
\end{equation}
In our experiments the confining stress is zero: $\sigma_2=\sigma_3=0$. Hence, from 
equation (\ref{eqn:qp}), at failure we have
\begin{equation}
\sigma_1 = \sigma^Y = 2 c \frac{\cos \varphi}{1- \sin \varphi}
\label{eqn:sigma1}
\end{equation}
Finally, the substitution of $c$ from equation \ref{eqn:c} yields
\begin{equation}
\sigma^Y = \frac{2}{3} n_b \langle \ell \rangle \frac{\sin \varphi}{1-\sin \varphi} \langle f^Y \rangle
\label{eqn:sigmaY}
\end{equation}

Equation (\ref{eqn:sigmaY}) shows that the compressive strength is proportional to the 
microscopic force threshold and to the number density of the bonds. 
Notice that $n_b$ is simply half the average number of bonds per particle 
divided by the free volume, i.e. the mean volume $V_p$ of a Voronoi cell 
surrounding the particle:
\begin{equation}
n_b = \frac{z}{2 V_p}
\label{eqn:nb}
\end{equation}
By definition, given the solid fraction $\phi$ of the sample, $V_p \phi $ is the 
mean volume of a particle. Thus, 
\begin{equation}
V_p = \frac{\pi \langle d^3 \rangle}{6\phi}
\label{eqn:Vp}
\end{equation} 
where $d$ is the particle diameter. 
For nearly spherical particles, we may 
set $\langle \ell \rangle \simeq \langle d \rangle$.  From equations (\ref{eqn:sigmaY}), (\ref{eqn:nb}) 
and (\ref{eqn:Vp}), we get the following equation in which the 
prefactor to $f^Y$ is expressed in terms of the familiar parameters:
\begin{equation}
\sigma^Y = \frac{2}{\pi} z \phi \frac{\sin \varphi}{1-\sin \varphi} \frac{\langle 
d \rangle}{\langle d^3 \rangle}  \langle f^Y \rangle
\label{eqn:sigmaY2}
\end{equation}
For the typical values $z=6$, $\varphi = \pi/6$ and $\phi=0.6$ and assuming monodisperse particles, 
we get $\sigma^Y \simeq 2 f^Y/ d^2$, which is a simple relation.

\subsection{Compressive strength}

From the relations (\ref{eqn:Icic}) and (\ref{eqn:sigmaY2}) together with the 
parametric adhesion law (\ref{eqn:loi_evolutive}), it is straightforward to 
predict the macroscopic behavior. We first take the average of both sides 
of equation (\ref{eqn:loi_evolutive}):
\begin{equation}
\langle f^Y \rangle = f^Y_L + \frac{f^Y_S - f^Y_L}{1 - i_T}  \langle (i_c-i_{T}) H \left( i_c - i_{T} \right)  \rangle
\label{eqn:fYave}
\end{equation} 
where the averaging runs over the whole sample. Inserting this expression of the mean force threshold 
in equation (\ref{eqn:sigmaY2}), we get
\begin{equation}
\sigma^Y = \frac{2}{\pi} z \phi \frac{\sin \varphi}{1-\sin \varphi} \frac{\langle 
d \rangle}{\langle d^3 \rangle} \left\{f^Y_L + \frac{f^Y_S - f^Y_L}{1 - i_T}  \langle (i_c-i_{T}) H \left( i_c - i_{T} \right)   \rangle \right\}
\label{eqn:sigmaY3}
\end{equation}

The average $ \langle (i_c-i_{T}) H \left( i_c - i_{T} \right) \rangle_V$ in the right hand side of 
equation (\ref{eqn:sigmaY3}) depends on the function $i_c(r,t)$ and will be evaluated below 
in the presence of a hydric gradient. Here, we consider the first and third regimes 
identified in the experiments (see section \ref{sec:experiments}). The first regime corresponds 
to the initial stage of drying where $i_c(r,t) < i_T$ for all bonds and no cemented 
bond is formed. Hence, according to (\ref{eqn:sigmaY3}), 
the compressive strength in this regime is given by
\begin{equation}
\sigma^Y = \sigma^Y_L = \frac{2}{\pi} z \phi \frac{\sin \varphi}{1-\sin \varphi} \frac{\langle 
d \rangle}{\langle d^3 \rangle} f^Y_L 
\label{eqn:sigmaYL}
\end{equation}    
We see that $\sigma^Y$ in this regime is constant and independent of $I_c$. The purely 
liquid-bond strength $\sigma^Y_L$ can be used to scale $\sigma^Y$ in 
equation (\ref{eqn:sigmaY3}):
\begin{equation}
\frac{\sigma^Y}{\sigma^Y_L} = 1 + \frac{{f^Y_S}/{f^Y_L}-1}{1 - i_T}  \langle  (i_c-i_{T}) H \left( i_c - i_{T} \right)  \rangle 
\label{eqn:sigmaY4}
\end{equation}
 
In the third regime, the cementation front has reached the center of the sample and 
we have $ i_T \leq i_c(r,t) $, and thus $H( i_c - i_T )=1$ everywhere. As a result, from 
equations (\ref{eqn:Icic}) and (\ref{eqn:sigmaY4}) we get
\begin{equation}
\frac{\sigma^Y_S}{\sigma^Y_L} = 1 + \frac{{f^Y_S}/{f^Y_L}-1}{1 - i_T}  (I_c-i_T)  
\label{eqn:sigmaY4.1}
\end{equation}
This equation shows that in the last stage of drying, 
corresponding to the consolidation of all bonds, the compressive strength increases 
linearly with $I_c$. 
This linear behavior and the predicted value of the coefficient $\frac{{f^Y_S}/{f^Y_L}-1}{1-i_{T} }\simeq 30$ for 
${f^Y_S}/{f^Y_L} \simeq 7$ and $i_T \simeq 0.9$  are 
consistent with the data shown in Fig. \ref{fig:sigmaYIc_exp} within 
our experimental precision.  

\subsection{Cementation front}
The second regime is a  transient beginning with the 
cementation of the bonds located at the surface and 
ending when all bonds are partially cemented. 
In this regime, the bonds at a radial position $r < r_T(t)$ are 
liquid whereas the bonds with $r_T(t) \leq r$ are partially cemented. 
The cementation front $r_T(t)$ decreases from $R$ at the end of the first regime 
to $0$ at the beginning of the third regime. By definition, we have 
$i_c(r=r_T,t=t_T)=i_T$. 

For the evaluation of the compressive strength in the second regime, we need 
the function $i_c(r,t)$. By definition, $1-i_c(r,t)$ is the proportion of the liquid brine 
in the bond, which is conserved in the bulk: The time variation $\partial_t\{1-i_c(r,t)\}$ 
of the liquid in the bonds 
located at $r$ is compensated by a flux $J$ in the vapor phase.  
Assuming that $J$ is proportional to the gradient $\partial_r \{1- i_c(r,t)\}$, we 
arrive at a diffusion equation for $i_c$:
\begin{equation} 
\frac{\partial i_c}{\partial t} = D \left\{ \frac{\partial^2 i_c}{\partial r^2} + \frac{1}{r} \frac{\partial i_c}{\partial r} \right\}
\label{eqn:diffusion}
\end{equation}
where invariance by rotation around the axis and by translation 
along the axis are assumed, and $D$ is a diffusion coefficient. 

Evaporation takes place only radially and at the 
free surface $r=R$. The exponential increase of $I_c$ observed in 
the experiments (see Fig. \ref{fig:Ict_exp}) 
suggests a simple exponential time evolution of $i_c(r=R,t)$:
\begin{equation}
i_c(R,t) = 1 - e^{-\beta t/t_T}
\label{eqn:icR}
\end{equation} 
The characteristic time $t_T$ represents the time necessary for 
the onset of cementation in a bond when $i_c=i_T$. 
We have $i_T=1-e^{-\beta}$. Hence, equation (\ref{eqn:icR}) can be rewritten 
in the following form     
\begin{equation}
i_c(R,t) = 1 - (1-i_T)^{t/t_T}
\label{eqn:icR2}
\end{equation} 

\begin{figure}
	\centerline{\includegraphics[width=.7\columnwidth]{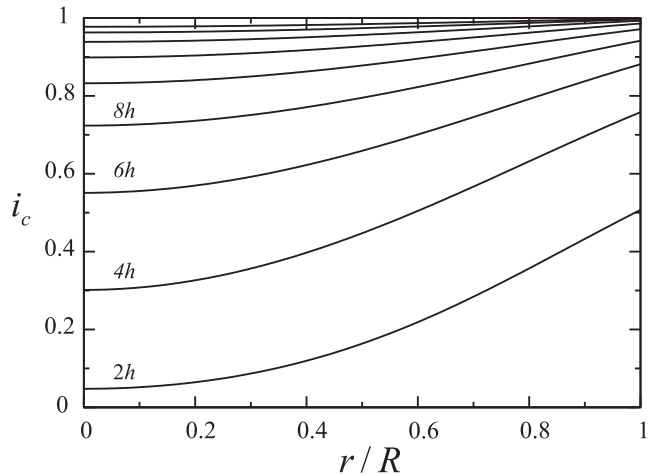}}
	\caption{Evolution of the local crystallisation index $i_c$ in the bulk with time.}
	\label{fig:ic_r}
\end{figure}

Equation (\ref{eqn:diffusion}) can be solved with (\ref{eqn:icR2}) as 
its boundary value for an initially homogeneous system in which $i_c(r,t=0)=0 $ for 
all $r \ne R$. The solution is \cite{Crank56}
\begin{equation}
i_c(r,t) = \{ 1 - (1-i_T)^{t/t_T} \}\left\{ 1- \frac{2}{R} 
\sum\limits_{n=1}^{\infty} \frac{e^{-D \alpha_n^2 t} J_0(\alpha_n r)}{\alpha_n J_1(\alpha_nR)}       \right\}
\label{eqn:Bessel}
\end{equation}
where $J_0$ and $J_1$ are Bessel functions of the first kind of order zero 
and one, respectively, and $\alpha_n$ is the $n^{\mbox{th}}$ root of the equation 
$\alpha J_1(\alpha R) = J_0(\alpha R)$. The solution (eq. \ref{eqn:Bessel}) is displayed
on  Fig. \ref{fig:ic_r}. 

From equation (\ref{eqn:Bessel}) we can calculate the global crystallization 
index defined by  (\ref{eqn:Icic}):
\begin{equation}
I_c (t)= \frac{2}{R^2} \int\limits_0^R  r i_c(r,t) \ dr =  \{ 1 - (1-i_T)^{t/t_T} \}\left\{ 1- \frac{4}{R^2} 
\sum\limits_{n=1}^{\infty} \frac{e^{-D \alpha_n^2 t}}{\alpha_n^2 }       \right\}
\label{eqn:Ic_Global}
\end{equation}
The macroscopic transition index $I_T$ is the value of $I_c$ at $t=t_T$. Hence
\begin{equation}
I_T= i_T \left\{ 1- \frac{4}{R^2} \sum\limits_{n=1}^{\infty} \frac{e^{-D \alpha_n^2 t_T}}{\alpha_n^2 }       \right\}
\label{eqn:IT}
\end{equation}
This relation implies that $I_T < i_T$. Given the high value of $t_T$, in 
practice we have  $I_T \simeq  i_T$.   

Equation (\ref{eqn:Bessel}) is an implicit equation for the cementation front $r_T(t)$ 
defined by the condition $i_c(r_T,t)=i_T$. The knowledge of $r_T(t)$ allows us  
to determine the total compressive strength $\sigma^Y$ as the 
weighted mean of  $\sigma^Y_L$ in the region $r<r_T$ and   $\sigma^Y_S$ 
in the region $r_T<r$:
\begin{equation} 
\sigma^Y(t) = \frac{r_T^2(t)}{R^2} \sigma^Y_L + \left\{1-\frac{r_T^2(t)}{R^2}\right\} \sigma^Y_S
\label{eqn:sigmaY6}
\end{equation}
Plugging the expressions (\ref{eqn:sigmaYL}) and (\ref{eqn:sigmaY4}) into equation (\ref{eqn:sigmaY6}) yields
\begin{equation}
\frac{\sigma^Y}{\sigma^Y_L} =   1+ \left\{1-\frac{r_T^2(t)}{R^2}\right\} \frac{{f^Y_S}/{f^Y_L}-1}{1 - i_T}  (I_c-i_T)  
\label{eqn:sigmaY7}
\end{equation}
where $i_T$ may be replaced by $I_T$ from equation \ref{eqn:IT}. 

Equation (\ref{eqn:sigmaY7}) holds only in the second regime, 
but we retrieve the expression of $\sigma^Y_S$ in equation (\ref{eqn:sigmaY4}) for 
the third regime by setting $r_T=0$ as well as the first regime 
by setting $r_T=R$ (equation (\ref{eqn:sigmaY4})). 
The nonlinear increase of  $\sigma^Y$ in the second regime with $I_c$ is given by the 
factor $1-{r_T^2}/{R^2}$. This corresponds to a first-order phase transition at 
the macroscopic scale for $I_c$ as control parameter and $\sigma^Y$ as state 
parameter.     

The model described in this section is able to predict 
the global crystallization index $I_c(t)$ as a function of time (equation (\ref{eqn:Ic})) and  
the compressive strength $\sigma^Y(I_c)$ as a function of $I_c$ 
(equations (\ref{eqn:sigmaY7}), (\ref{eqn:sigmaY4}) and (\ref{eqn:sigmaY4})). 
The plots of these functions for the experimental parameters are displayed in 
figures \ref{fig:Ict_exp} and \ref{fig:sigmaYIc_exp}. The predicted behavior is thus fairly good  
within the available experimental precision. Apart from the radius $R$ of the 
sample, for these fits three parameters were used from experimental measurements, 
namely $i_T$, $D$ and $t_T$. The good estimation of $\sigma^Y_L$ 
provides a strong support for our model.     


\begin{figure}
	\centerline{\includegraphics[width=.7\columnwidth]{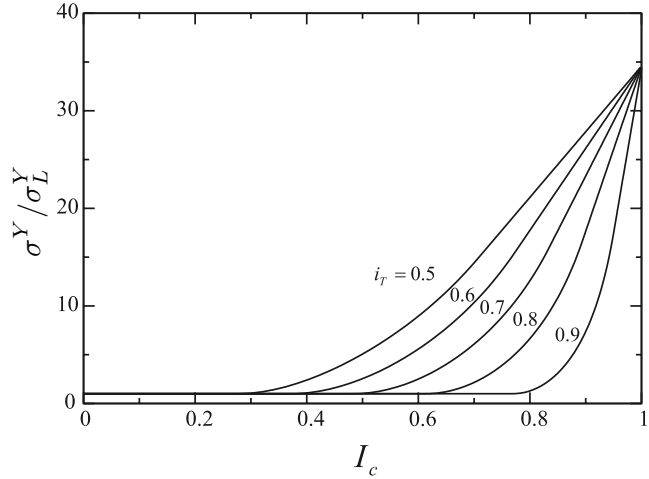}}
	\caption{Influence of the local transition index on the compression strength as a function
	of the macroscopic crystallisation index $I_c$.}
	\label{fig:}
\end{figure}

\section{Conclusion}

In this paper, we introduced a simple experimental setup allowing 
us to analyze the transition from capillary cohesion to solid cohesion 
induced by surface drying and the progressive 
crystallization of a liquid brine wetting initially the particles.  
This transition was evidenced experimentally by measuring the 
compressive strength of the samples in  the course of time, and modeled 
by introducing the relevant variables both at the scale of a bond and 
at the macroscopic scale. Three regimes were evidenced: 
1) the capillary regime as long as the degree of saturation remains below 
a critical value; 2)  the transition regime, corresponding to the initiation and propagation 
of a cementation front spreading towards the center of the sample, 
and  3) the cemented regime, where all bonds are partially solidified  
but continue to strengthen due to further crystallization. 
The predicted behavior by our theoretical model is in good agreement with the 
experimental data. This model, based on an evolutive cohesion law at the bonds, 
takes into account the kinetics of crystallization and the granular texture. 
It correctly predicts the compressive strength in the capillary and 
cemented regimes.      

The framework of the model presented in this work is quite general. 
It can be applied to granular materials subjected to evolutive force 
laws or phase compositions. For example, the degradation of a geomaterial 
under the action of progressive dissolution of solid bonds in a humid 
environment  bears strong similarity with the problem treated in this paper. 


\clearpage

\bibliographystyle{plain}

\end{document}